\documentclass[12pt,preprint]{aastex} 

\shorttitle{Ultraviolet Spectra of ULXs}
\shortauthors{Bregman et al.}

\begin{document}


\title{Ultraviolet Spectra of ULX Systems}


\author{Joel N. Bregman, Julie N. Felberg, Patrick J. Seitzer}
\affil{Department of Astronomy, University of Michigan, Ann Arbor, MI  48109}
\email{jbregman@umich.edu}

\author{Jifeng Liu}
\affil{Eureka Scientific, Oakland, CA 94602}

\and

\author{Martin K\"{u}mmel}
\affil{Space Telescope Science Institute, Baltimore, MD 21218}


\begin{abstract}
To further understand the nature of the optical counterparts associated with
Ultraluminous X-Ray sources (ULXs), we obtained far ultraviolet spectra of the
reported counterparts to the ULXs in NGC 1313, Holmberg II, NGC 5204, and M81.
The spectral resolution of the ACS prism spectra degrades from 300 at 1300\AA\
to 40 at 1850\AA\, so longer wavelength features have lower S/N.  The spectra
of the ULXs in NGC 1313, Ho II, and NGC 5204 are quite similar, showing the N V
$\lambda$1240 line at about the same equivalent width strength.  The presence
of this and other emission lines confirms the presence of an accretion disk,
probably diluted by the light of an early B star companion.  The spectra differ
strongly from high mass X-ray binaries dominated by O star winds (e.g., Cyg
X-1) and are most similar to intermediate mass X-ray binary systems in the Milky
Way and LMC.  This indicates that the mass transfer is due to Roche Lobe overflow.
The spectrum of the ULX in M81 is quite weak but suggestive of a late-type 
Wolf-Rayet star.

\end{abstract}


\keywords{X-rays: binaries }


\section{Introduction}

Ultraluminous X-ray sources (ULXs) are a class of X-ray point sources, mostly
in the disks of spiral galaxies, with X-ray luminosities exceeding the
Eddington luminosity (L$_{Ed}$)of the most massive known stellar mass black
holes in the Milky Way, Cygnus X-1 (16 M$\odot$ and L$_{Ed}$ =
2$\times$10$^{39}$ erg s$^{-1}$; Fabbiano 2006).  There is typically one ULX
per spiral galaxy with luminosities sometimes exceeding 10$^{40}$ erg s$^{-1}$
(e.g., Liu \& Bregman 2012).  A few explanations are generally offered to
explain ULXs, all involving mass transfer from a donor star to a black hole.
These ULXs could be intermediate mass black holes (10$^2$-10$^4$ M$_\odot$)
that are emitting at 0.01-0.1 of L$_{Ed}$, similar in efficiency to their
lower-mass counterparts (Colbert and Mushotzky 1999).  Another explanation is
that they are exceeding  L$_{Ed}$ due to the accretion disk somehow
accommodating this higher luminosity through exotic non-axisymmetric processes,
such as photon bubbles (Begelman 2002).  A third possibility is that the
emission is highly anisotropic (King et al. 2001).

Not only is there uncertainty regarding the mass of the primary, there are
several important unresolved issues regarding the nature of these binary
systems, which, if resolved, could help appreciate the underlying physical
situation.  One of the issues is the nature of the secondary that is providing
the mass transfer.  It was expected that the secondary would be massive, as
ULXs are young phenomena, primarily lying in star-forming regions that are
typically 10 Myr or less.  They are not in extremely dense star clusters, so
dynamical interactions with other stars is unlikely in the lifetime of the ULX
(we can treat them as isolated binaries).  We anticipated that the secondary
would have begun evolving away from the main sequence, so it would have the
magnitude of an O star, but it could have any color as it expands to fill its
Roche Lobe.  However, when optical counterparts were discovered (mainly with
HST), they were found to be blue, even after correcting for possible
contamination by an accretion disk (e.g., Liu et al. 2002).  This leads to the
conjecture that the mass transfer is related to massive stellar winds from hot
stars.  

This conjecture can be tested with ultraviolet spectra of these systems, as
hot stars can have distinctive far ultraviolet features.  Such a program will
address two issues, the first being the identification of a particular stellar
object with the ULX.  Many optical identifications depend on the proximity of
an optical object within the error circle of an X-ray identification.  Some of
these identifications may occur by chance because ULXs occur in moderately
crowded fields.  Real and false identifications can be distinguished by UV
spectra, as X-ray binaries have distinctive emission lines.  The other goal is
to constrain the spectral type and properties of the secondary and of the
accretion disk.  This is easier to accomplish in the far UV because most hot
stars depart from the Rayleigh-Jeans limit (evident in the optical region) and
the strongest absorption lines are from high excitation ions, which occur in
the UV region.

We study the far UV prism spectra, obtained with the Advanced Camera for
Surveys on HST, for four ULXs in this paper.  We first describe the target
selection and data processing in \S 2, the analysis of the spectra in \S 3,
then the interpretation of these spectra in the context of spectra of known
objects and a color-color diagram in \S 4, followed by a discussion in \S 5. 

\section{Target Selection and Observations}

From the combination of Chandra and HST imaging, a number of optical
counterparts had been identified at the time this program began.  In order to
obtain spectra of these counterparts with a moderate number of orbits, they had
to be sufficiently bright in the ultraviolet region, which required that the
targets be nearby (D \textless 10 Mpc), with moderately low dust extinction,
and sufficiently high optical flux.  Another consideration was crowding, as ULX
optical counterparts can occur in complex regions.  With the narrow slit of the
STIS instrument, most confusion problems could be avoided, but when STIS
suffered a failure, the only viable UV spectrograph was the prism on the
Advanced Camera for Surveys.  However, as there are no slits on this
instrument, it could only be used for ULX fields in which source confusion
could be avoided.  After considering possible candidates and a variety of roll
angles, we found it was possible to obtain UV spectra of four ULX optical
counterparts in the galaxies Ho II, M81, NGC 1313, and NGC 5204 (Table~1).  The
data were obtained in Cycle 15.

For the observations, we used the Solar Blind Camera (SBC), in conjunction with
the PR130L prism, which covers the wavelength range 1230\AA\ to 1850\AA; the
observations comprised most of two orbits per object.  A short complementary
image was also obtained with the SBC, with the F165L filter, which allows one
to register the spectra with the point sources and to calibrate the wavelength
range.  To extract the spectra, we used the aXe software package (K\"{u}mmel et
al. 2009), following standard procedures.  The highest quality spectra were for
the ULX optical counterparts in NGC 1313 and Ho II, followed by NGC 5204, and
with M81 having the poorest spectrum due to its lower flux level.

\section{Analysis and Modeling of the Data}

For spectra obtained with dispersion gratings, there is only a modest variation
in the spectral resolution, while for prism spectra, the resolution can vary by
a factor of several, as is the case here.  Over the effective waveband of the
prism spectrum, 1220$-$1850\AA, the resolution decreases from about 300 at
1300\AA\ (1000 km s$^{-1}$) to about 100 at 1500\AA\ (3000 km s$^{-1}$), to
40 at 1850\AA\ (7500 km s$^{-1}$).  Consequently, it is easier to detect lines
and their shapes at the short wavelength end, as we see in Figure 1.  

\begin{figure}

\plottwo{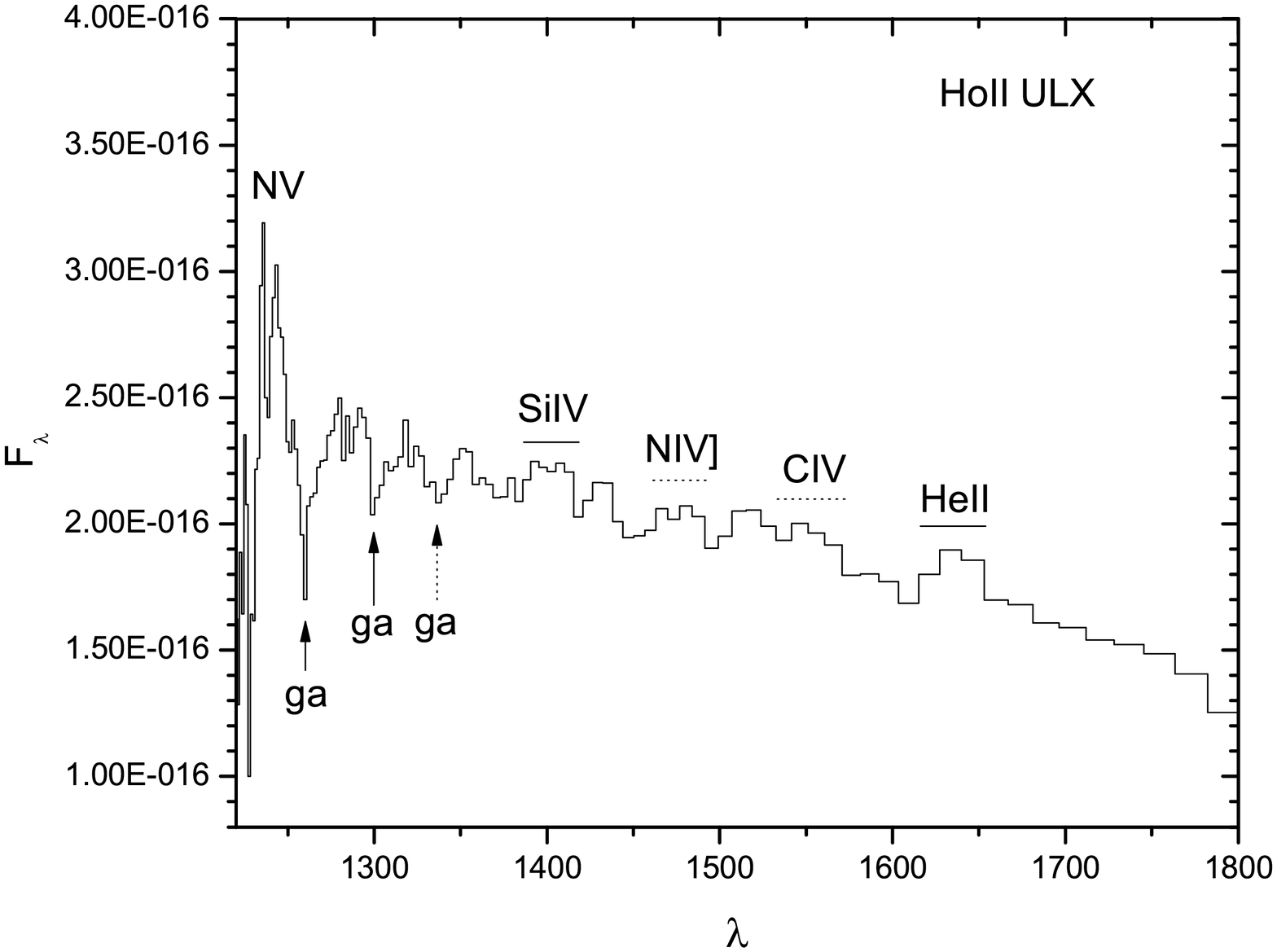}{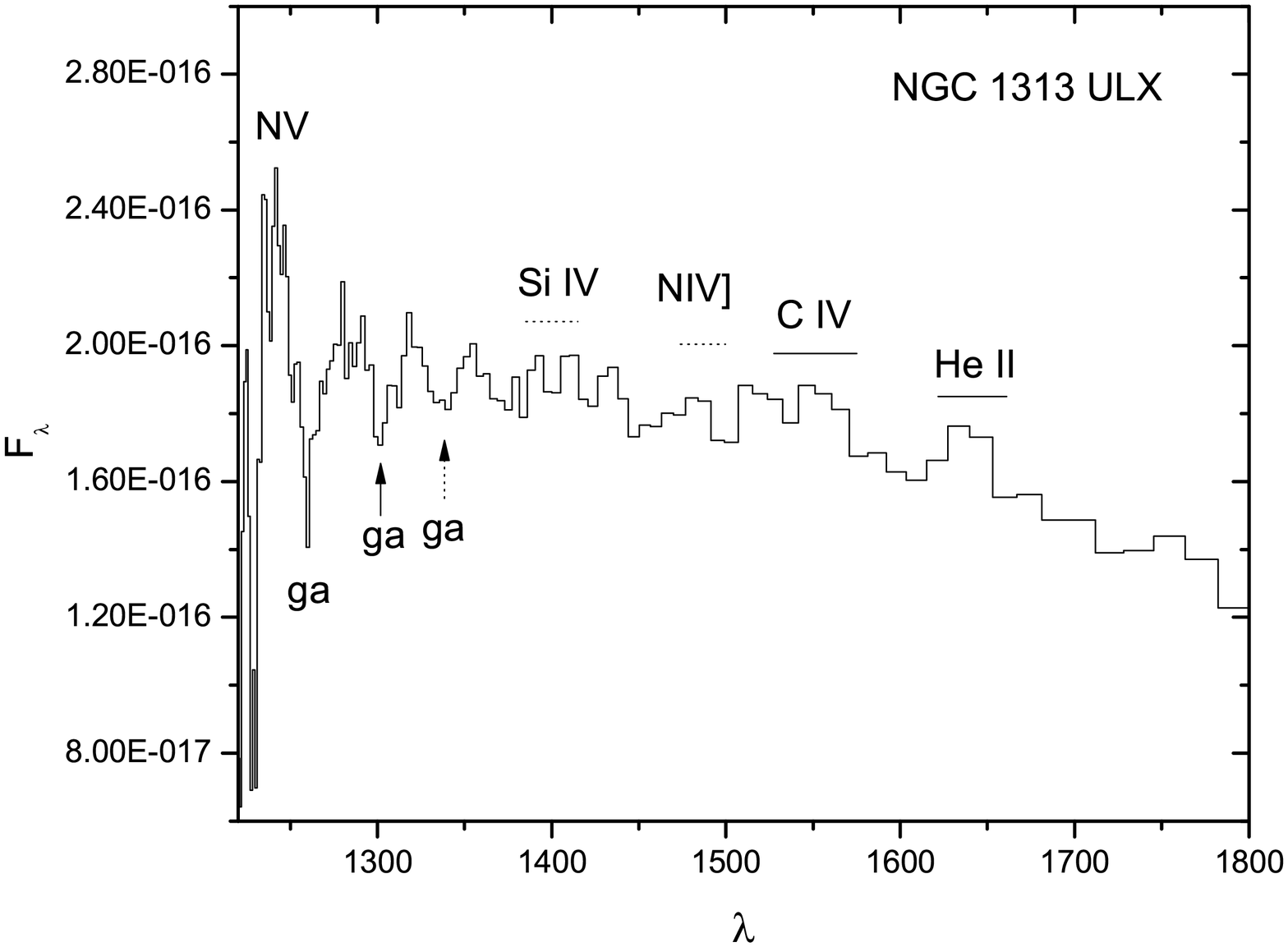}

\plottwo{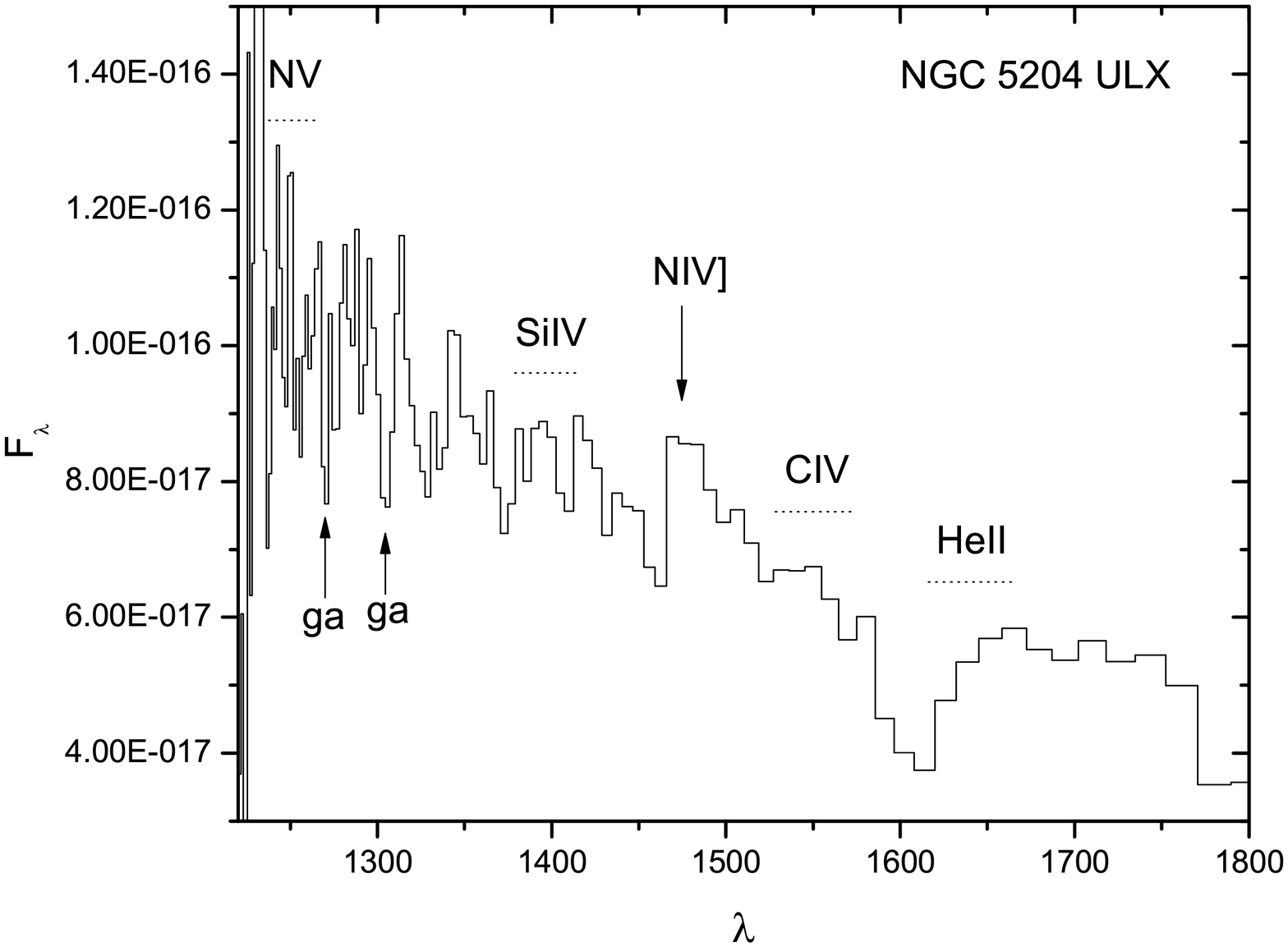}{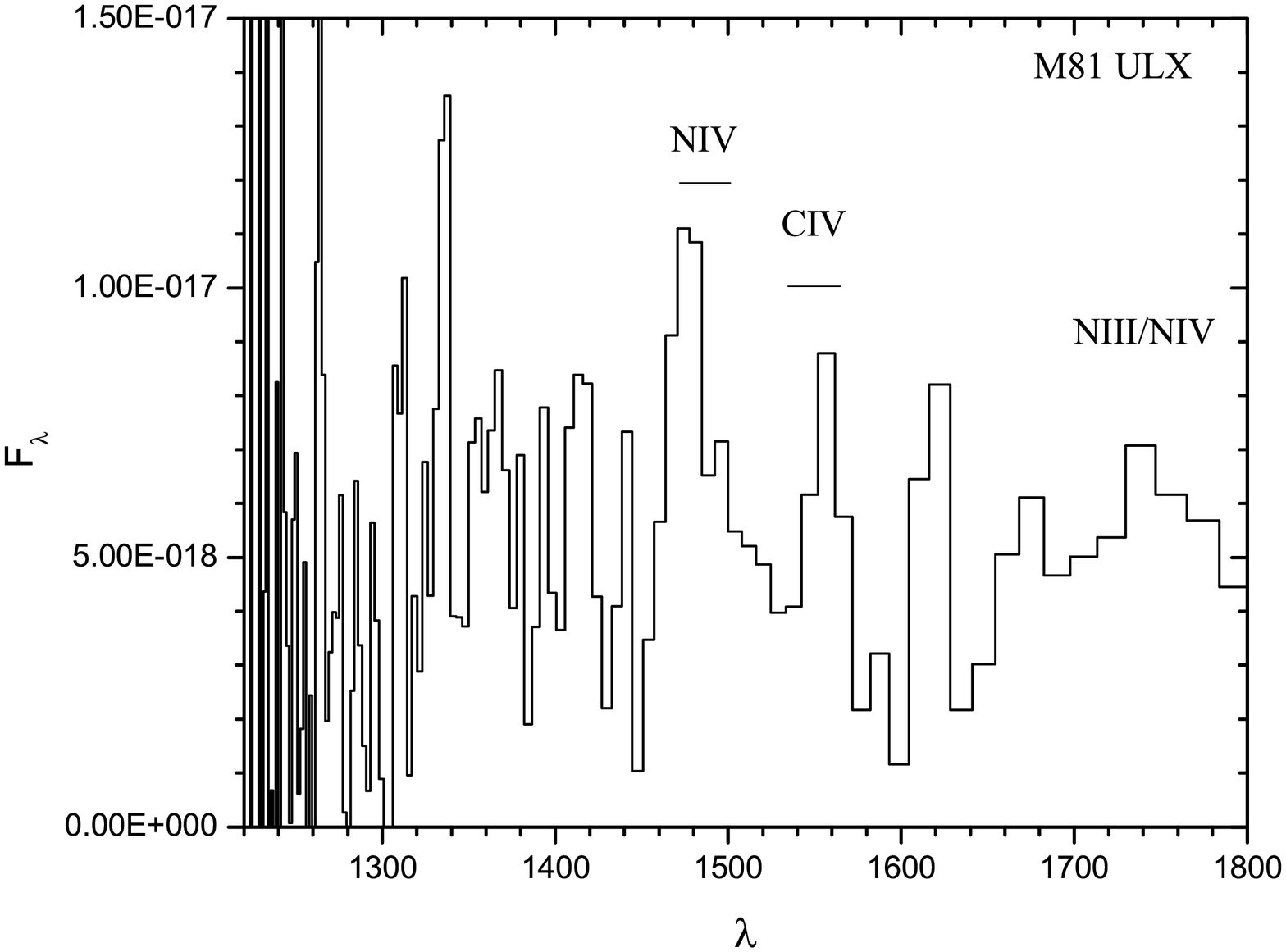}

\caption{The HST/ACS prism spectra for the four ULXs in (a) Ho II, (b) NGC1313,
(c) NGC5204, and (d) NGC3031.  Labeled are the three galactic absorption lines
(Si II $\lambda$1260.4, O I $\lambda$1302.2, and C II $\lambda$1334.5), doublet
lines NV $\lambda$1238.8, 1242.8, Si IV $\lambda$1393.8, 1402.8, C IV
$\lambda$1548.2, 1550.8, semi-forbidden line N IV $\lambda$1486.5, and He II
$\lambda$1640.4. Dotted lines under the symbol for the ion shows where the ion
should lie but that there is not a likely detection.  Talactic absorption ines 
are noted with the symbol ga. The range in the flux density is similar, a factor of 
five, except for (d).}

\end{figure}

We sought to identify known UV lines in these ULX systems.  These systems have
several components, including an accretion disk and a hot stellar companion, and
where the spectrum is modified by Galactic absorption.  The lines anticipated from
each component can be identified, making a decomposition possible.  \\
$\indent$ Galactic absorption lines: the strongest that would be detected in
this wavelength region are Si II $\lambda$1260.4, O I $\lambda$1302.2, and C II
$\lambda$1334.5.  These lines are completely unresolved, so they provide a
local measurement of the line width.  \\
$\indent$ Accretion Disks in X-Ray Binaries: Emission lines that are commonly
found in X-ray binaries with accretion disks include the following high ionization
lines:  NV $\lambda$1238.8, 1242.8, Si IV $\lambda$1393.8, 1402.8, C
IV $\lambda$1548.2, 1550.8, and He II $\lambda$1640.4. \\
$\indent$ OB Stars:  Photospheric absorption lines that we searched for are Si
II $\lambda$1264.7, Si III $\lambda$1298.9, O IV $\lambda$1338.6, and Si III
$\lambda$1417.2 (Prinja 1990).  These lines are not resonance lines, so the
lower level is populated through frequent collisions, which occur in the
atmosphere of the star.  There are other stronger lines in this wavelength
range (e.g., He II), but they cannot be detected with the resolution of these
prism data.  P$-$Cygni emission lines are sometimes seen in OB stars due to
their winds, primarily in the doublets of C IV $\lambda$1548.2, 1550.8, and Si
IV $\lambda$1393.8 (usually weaker). \\
$\indent$ Wolf-Rayet Stars: These stars have winds with high rates of mass loss
and emission lines.  The strongest emission line is often He II
$\lambda$1640.4, but the other prominent lines depend on whether the star has a
relative excess of nitrogen or carbon, being the WN and WC designations
(Niedzielski and Rochowicz 1994).  For the WN stars, the N IV $\lambda$1486.5
line can be prominent and it is distinctive in that it is associated with
high-density winds and is not predicted to be produced in accretion disks.  The
C IV $\lambda$1548.2, 1550.8 doublet is often detected in WN stars, with a
P$-$Cygni profile, although in the WC stars, this is usually the strongest
line, followed by the He II line.  There is a wide range in the UV spectrum of
WR stars, from having very prominent emission lines to rather weak emission.

	Our inspection of the four targets led to the identification of the
following features.  The results are given in order of decreasing continuum
flux, and therefore decreasing S/N.

Holmberg II: All three Galactic absorption features are detected.  In emission,
we detect the N V 1240 doublet and the He II 1640 line, and probably the Si IV
doublet as well.  Other features are not statistically significant, such as the
N IV 1486 line or the C IV 1550 doublet.  There is a hint of stellar absorption
features, which would need confirmation with higher resolution spectra.  The
stellar Si II $\lambda$1264.7 line lies just to the red of the Galactic Si II
$\lambda$1260.4 line, which is not symmetric on the red side in the sense that
it is consistent with an additional absorption feature that is typical of this
stellar line.  The next stellar line, at Si III $\lambda$1298.9, lies to the
blue of the Galactic O I $\lambda$1302.2 line.  The minimum of the absorption
feature corresponds to the Si III line rather than the O I line.  The next
stellar line, O IV $\lambda$1338.61, is also close to a Galactic absorption
line, C II $\lambda$1334.5, with the spectral minimum lying between the two
features.

We can infer the velocity width of the N V $\lambda$1240 emission lines by
comparing it to the Si II $\lambda$1260.4 line, which has a FWHM of about
3.6\AA, consistent with the instrumental resolution.  This is narrower than the
width of the N V $\lambda$1240 doublet, where the blue side is poorly defined
due to the rapidly decreasing sensitivity.  When we use the red side of this
emission line to define the line width, we obtain a FWHM of 6\AA.  The inferred
intrinsic FWHM is 5\AA, after compensating for instrumental resolution, so if
the FWHM is characteristic of the outflow or rotational velocity, it would be
about 600 km s$^{-1}$.  

NGC 1313: The spectrum is similar to that of Holmberg II, showing the same
Galactic absorption features, along with the N V and He II emission features.
There is likely broad emission corresponding to C IV $\lambda$1550, but other
features are not statistically significant.  There is a broadened red wing to
the Si II $\lambda$1260.4 line, which could be due to the stellar Si II
$\lambda$1264.7 line, but better data are needed to confirm this.  There are no
suggestions of the other stellar features.  The line width of the N V
$\lambda$1240 emission is the same, to within statistical uncertainties, with
the one seen in Holmberg II. 

NGC 5204: The continuum is a few times weaker than Ho II, so some lines visible
in Ho II would not be expected to be significant here.  The strongest Galactic
absorption line, Si II $\lambda$1260.4, is not detected.  The Galactic O I
$\lambda$1302.2 may have been detected, although the line center is at
1304.2\AA.  This shift could be due to absorption within the disk of NGC 5204.
In emission, the N V 1240 feature is not detected, nor is He II 1640 or C IV
1550.  There is an apparent emission feature with a peak at 1478.3\AA, so it is
possible that this is the WN feature N IV 1486, if there is a calibration
wavelength offset.  A local minimum at 1610\AA\ does not correspond to any
known feature.

Messier 81: The UV fluxes are about a factor of 20 lower than for Ho II, so
only emission lines would be evident, as a continuum would be too weak to
reliably detect.  The strongest feature has a central wavelength of about
1477\AA, as seen in NGC 5204.  This may be the WN feature N IV 1486, with a
wavelength offset.  There may be emission from the C IV 1550, which is close to
the correct wavelength (1556\AA), and the effect of a P$-$Cygni profile, if
present, is to shift the line to longer wavelength.  There is a weak peak at
the location of Si IV 1400.  There is no detection of He II 1640 emission.  Two
features at 1310\AA\ and 1337\AA\ are unidentified and are narrower than the
instrumental resolution, so unlikely to be real.

\section{Interpretation of the Spectra}

In order to understand the features and nature of the spectra, we took a
variety of UV spectra that represent different types of objects and convolved
them with the varying spectral resolution of the prism.  A brief discussion of
each class, with examples are given below and we have tried to retain the same
scale as for the ULX sources, which is a factor of five in the vertical scale
and for a wavelength range of 1220-1800\AA.

OB Stars: Most of the OB stars examined are dominated by a stellar continuum
with a few absorption features due to the intervening interstellar medium (Fig.
2a) as well as some weak stellar absorption line features, such as Si III
$\lambda$ 1298.9, Si III $\lambda$ 1417.2, or He II $\lambda$ 1640.4 (Prinja et
al. 1990).  A very hot star, such as a planetary nebula central star (Fig. 2b)
has nearly a Rayleigh-Jeans power-law shape over this wavelength range,
punctuated by absorbing features in the ISM.  Some O stars have stellar winds,
displaying P$-$Cygni profiles, signatures that we see strongly in the
Wolf-Rayet stars and the high mass X-ray binaries.

Wolf-Rayet Stars: There is a very wide range of spectral behavior for the
various classes of WR stars, but line emission is a common property, although
the line strengths can vary dramatically.  The WN4 (Fig. 2c) star shows a
typical spectrum, with a strong He II line, followed by lines of N V, Fe V/O V,
C IV, and N III/N IV.  The P$-$Cygni profiles are washed out at the resolution
of the PR130L and the peak of the line appears to shift from its normal value.
At prism resolution, the P$-$Cygni profile for C IV $\lambda$ 1550 results in
an absorption dip surrounded by local peaks, which is common for hot wind
outflows.  For a WN8 star (Fig. 2b), which has a lower degree of ionization,
the He II line is weaker, and at the resolution of the prism, it is not
significant.  Lines of NIII and NIV are visible and there is a dip at the
location of C IV $\lambda$ 1550 but it might not be considered significant.
There is no NV $\lambda$ 1240 emission and the blue part of the spectrum is
suppressed, either due to extinction or to a lower temperature for the star.
Toward earlier WN type (higher ionization), the He II $\lambda$ 1640 line
becomes dominant and the N V $\lambda$ 1240 line is present.  The WC stars have
a much stronger (and often dominant) C IV $\lambda$ 1550 line and weak nitrogen
lines.

\begin{figure}

\plottwo{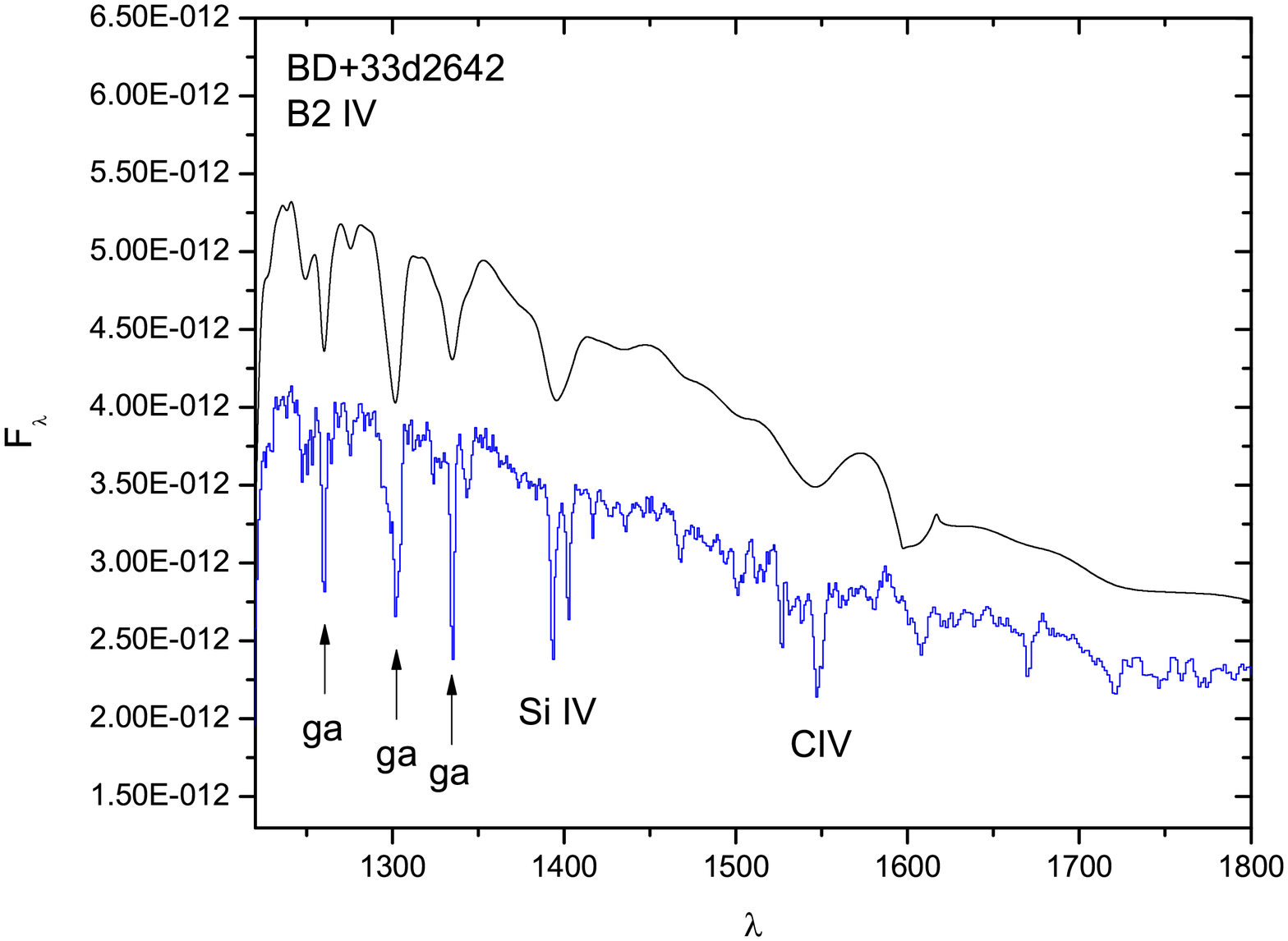}{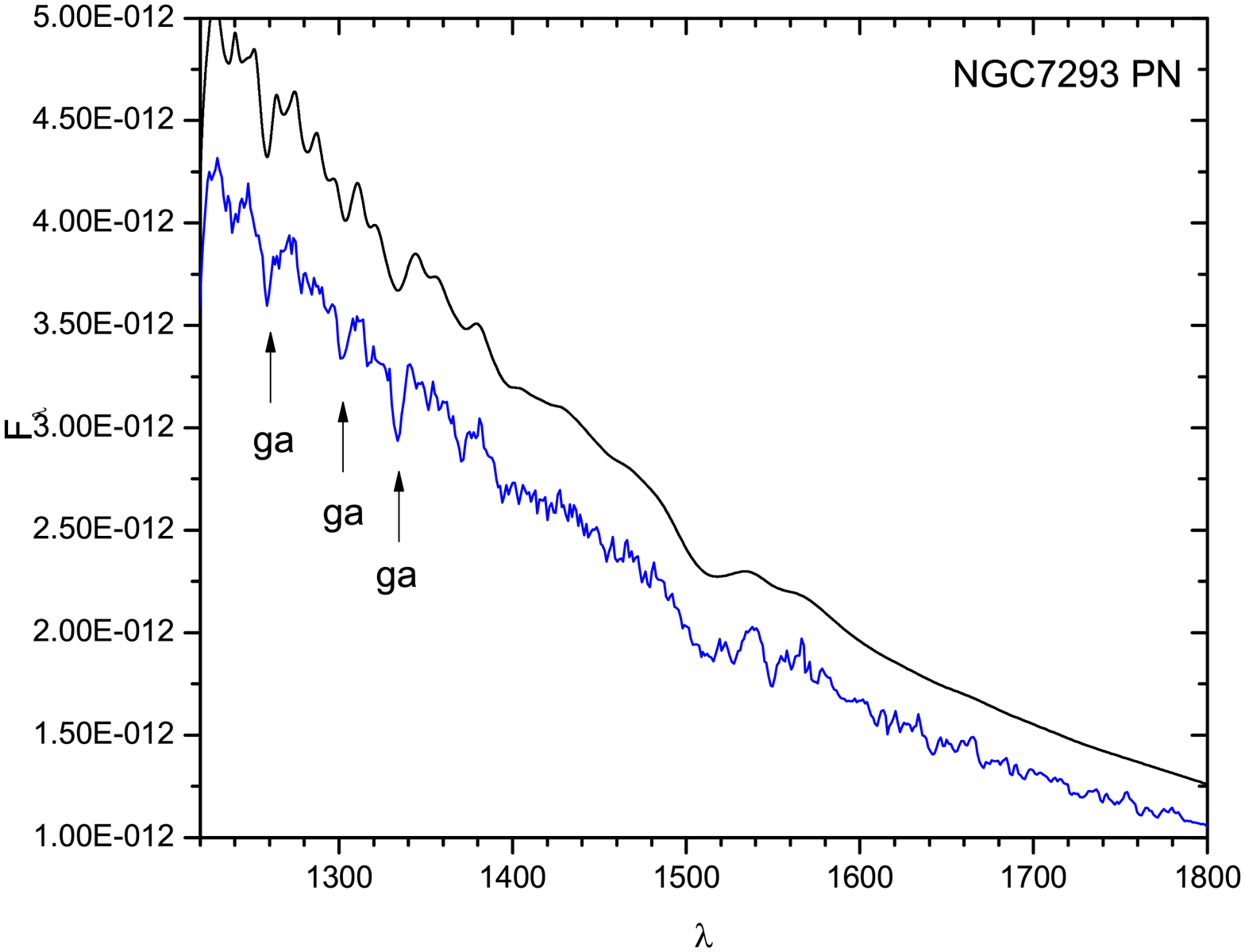}

\plottwo{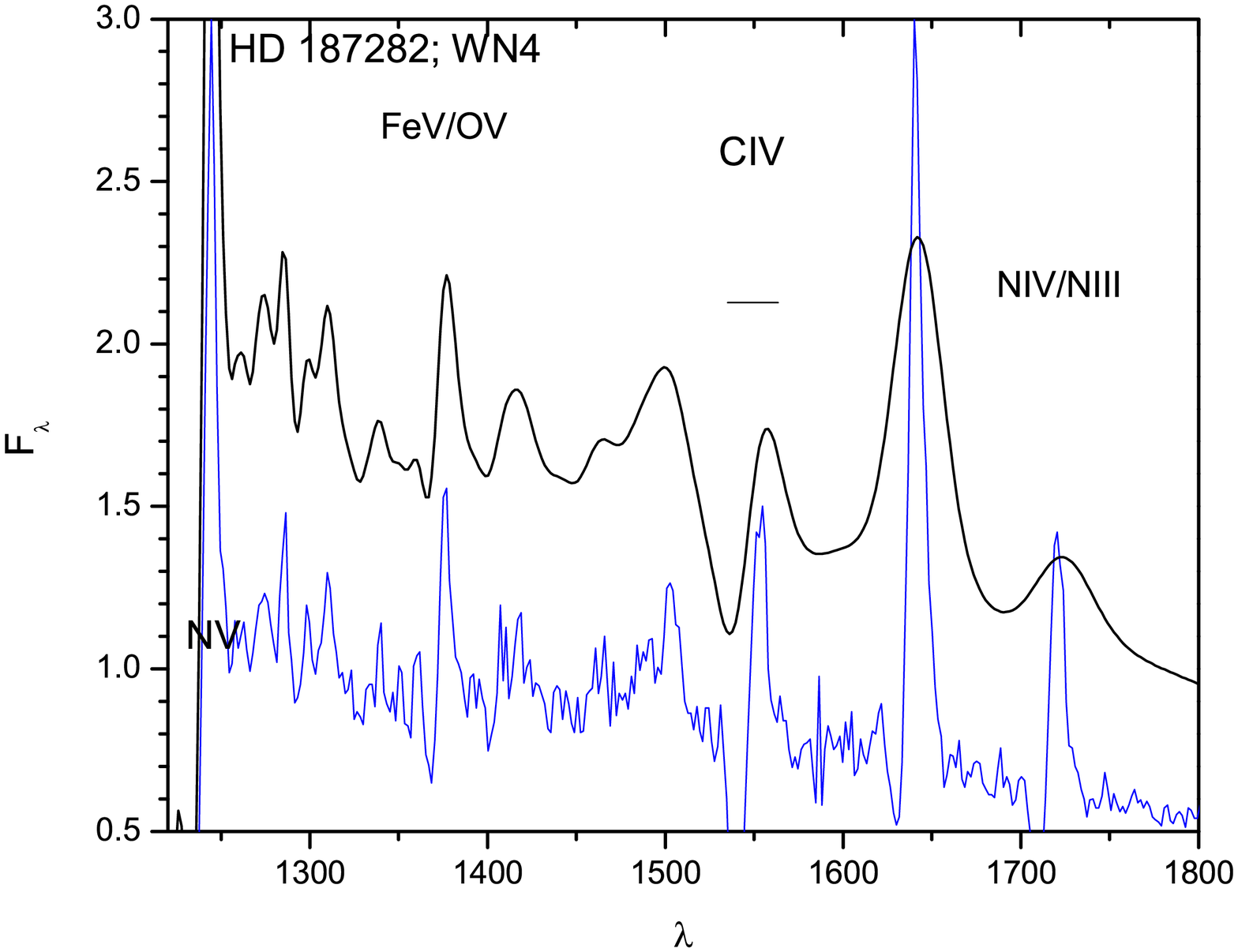}{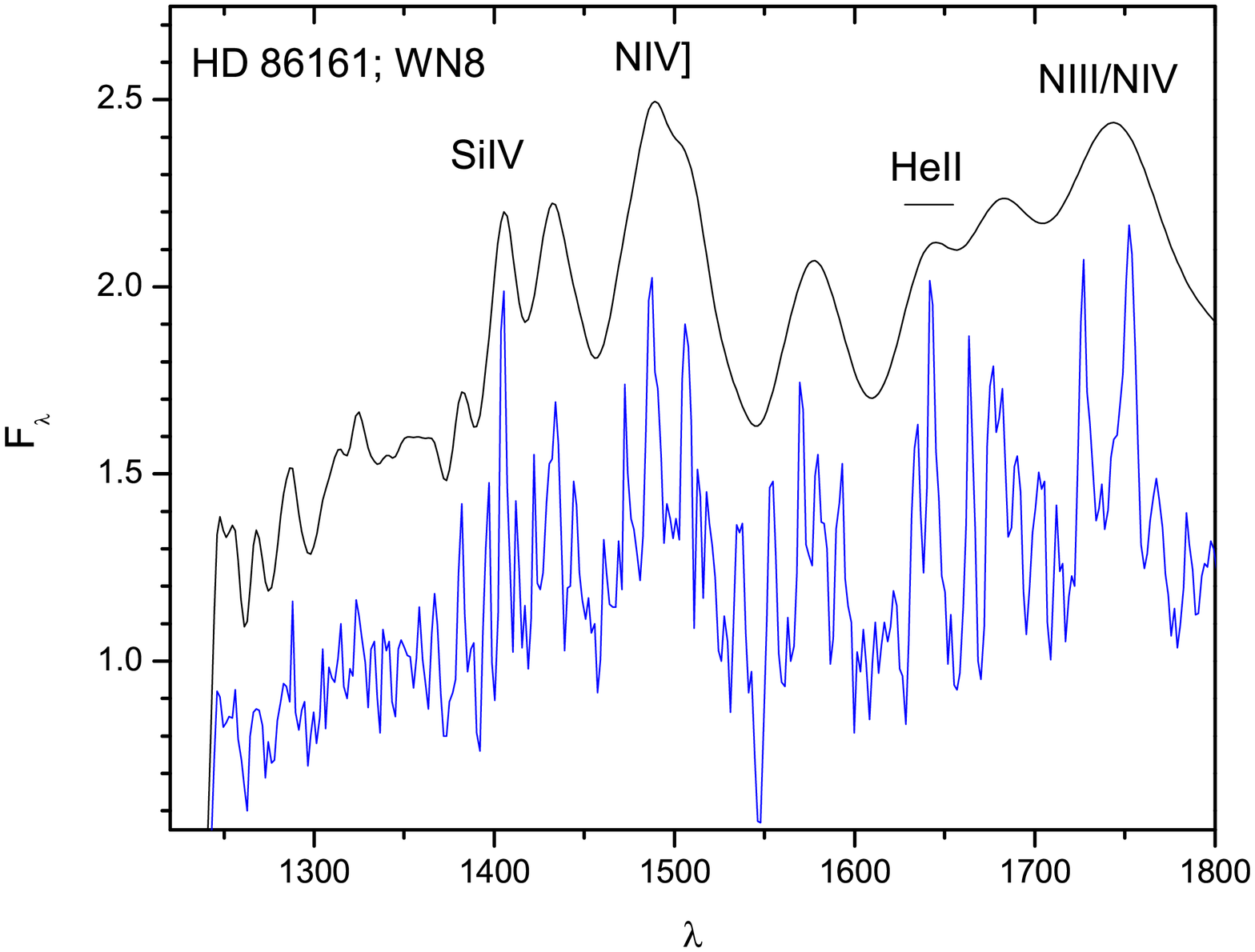}

\caption{The IUE spectra for (a) a hot B2IV star BD+33d2642,  (b) a planetary nebula
central star in NGC7293, (c) a WN4 star HD 187282, and (d) a WN8 star HD 86161.
The smoothed curves are obtained by convolving the IUE spectrum with the
varying spectral resolution of the HST/ACS prism. The range in flux density is 
a factor of five for all spectra except for (c), where it is a factor of six.}

\end{figure}

High Mass X-Ray Binaries: In these cases, the compact object is a black hole or
neutron star and the star is of early type with mass loss that includes and can
be dominated by a stellar wind.  The object 4U1700-37 (Fig. 3a) is a good
example where the compact object has a mass of 2.4 M$_\sun$, and it is not
clear if this is a NS or BH.  The massive star is type O6.5Iaf, with a mass of
58 M$_\sun$ and an orbital period  3.4 days.  It has strong P-Cygni
emission lines of NV, Si IV, and CIV, with a weak He II $\lambda$ 1640 line
(Fig. 3a).  At the resolution of the PR130L, only an absorption feature
survives near 1550\AA, the He II line is not significant, but the Si IV and NV
emission is present, with the P-Cygni profile still visible for Si IV.  Another
example of this class, Cyg X-1 (Fig. 3b) does not show P-Cygni profiles, but
does show absorption features corresponding to Si IV and C IV.

Intermediate Mass X-Ray Binaries: These systems have a black hole or neutron
star with a star of 2-6 M$_{\sun}$ in orbit.  Examples are LMX X-3 (black hole
plus 5.9 M$_{\sun}$ star) or Her X-1 (HZ Her; a neutron star plus a 2.3
M$_{\sun}$ star), and in the latter case, the strongest emission feature is NV
$\lambda$ 1240, followed by C IV $\lambda$ 1550 and He II $\lambda$ 1640, which
is a small feature at the resolution of the prism (Fig. 3c).  A N IV]
$\lambda$ 1486\AA\ line is visible in LMC X-3, along with C IV $\lambda$ 1550,
but the NV line is the strongest feature (Fig. 3d).

\begin{figure}

\plottwo{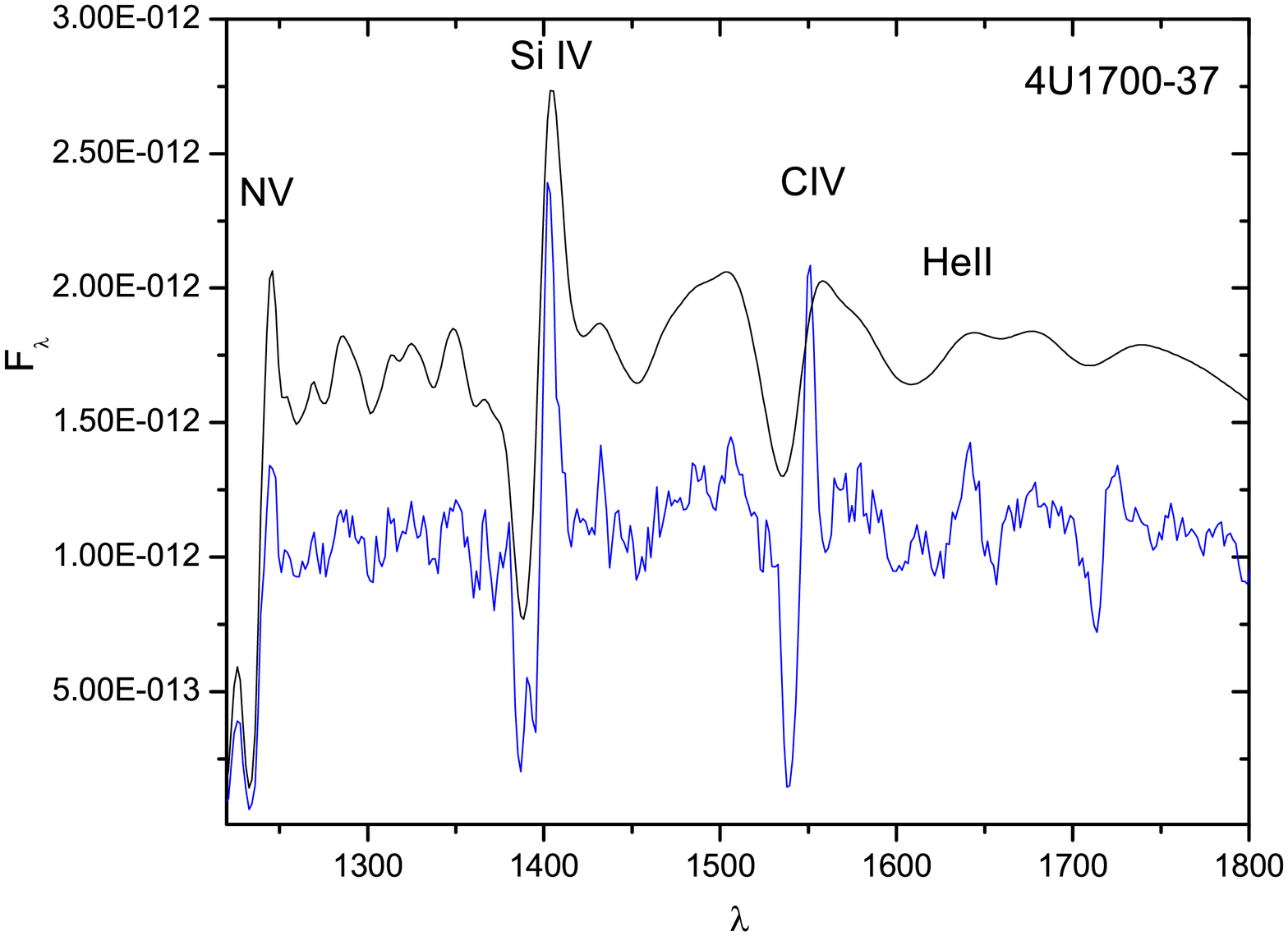}{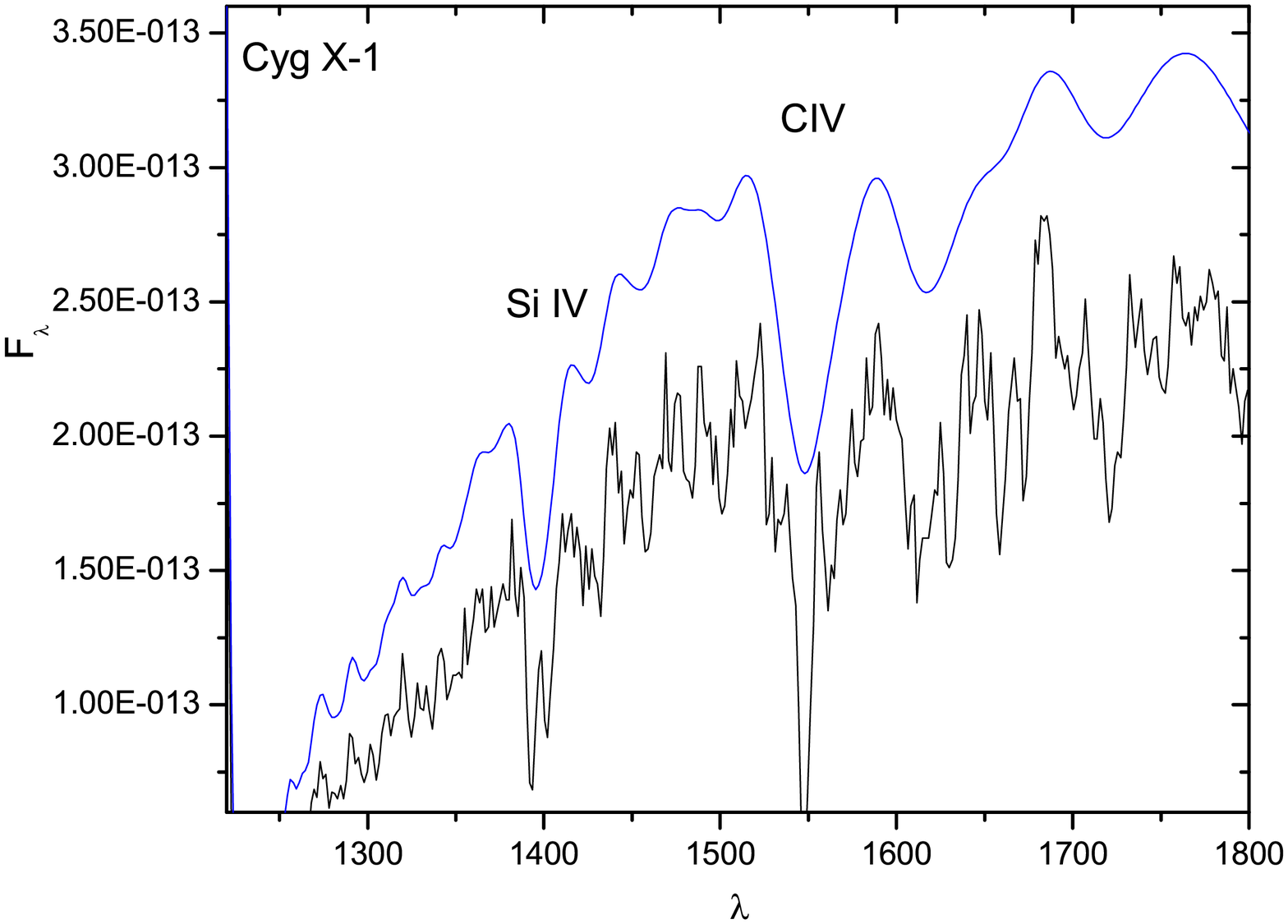}

\plottwo{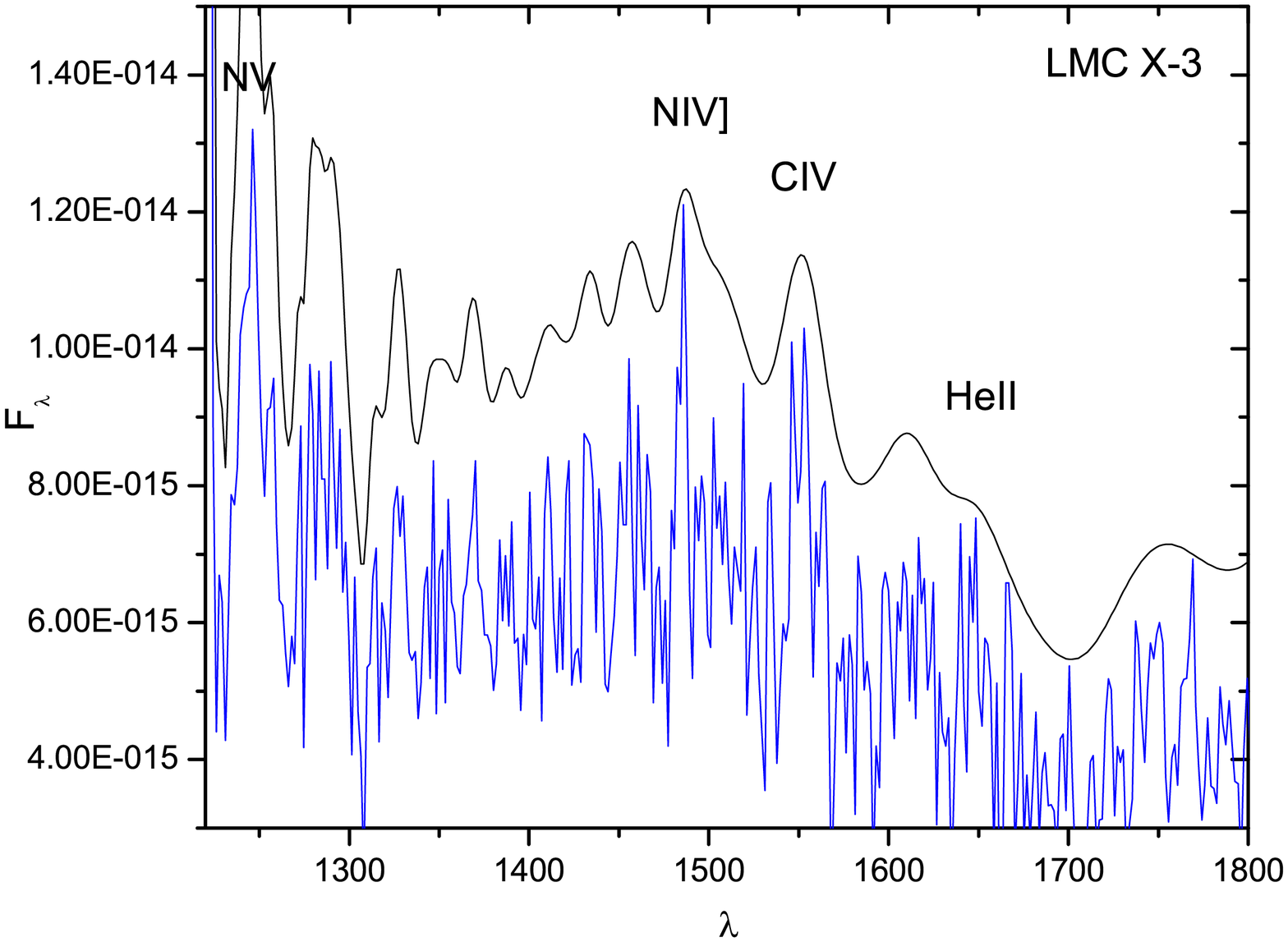}{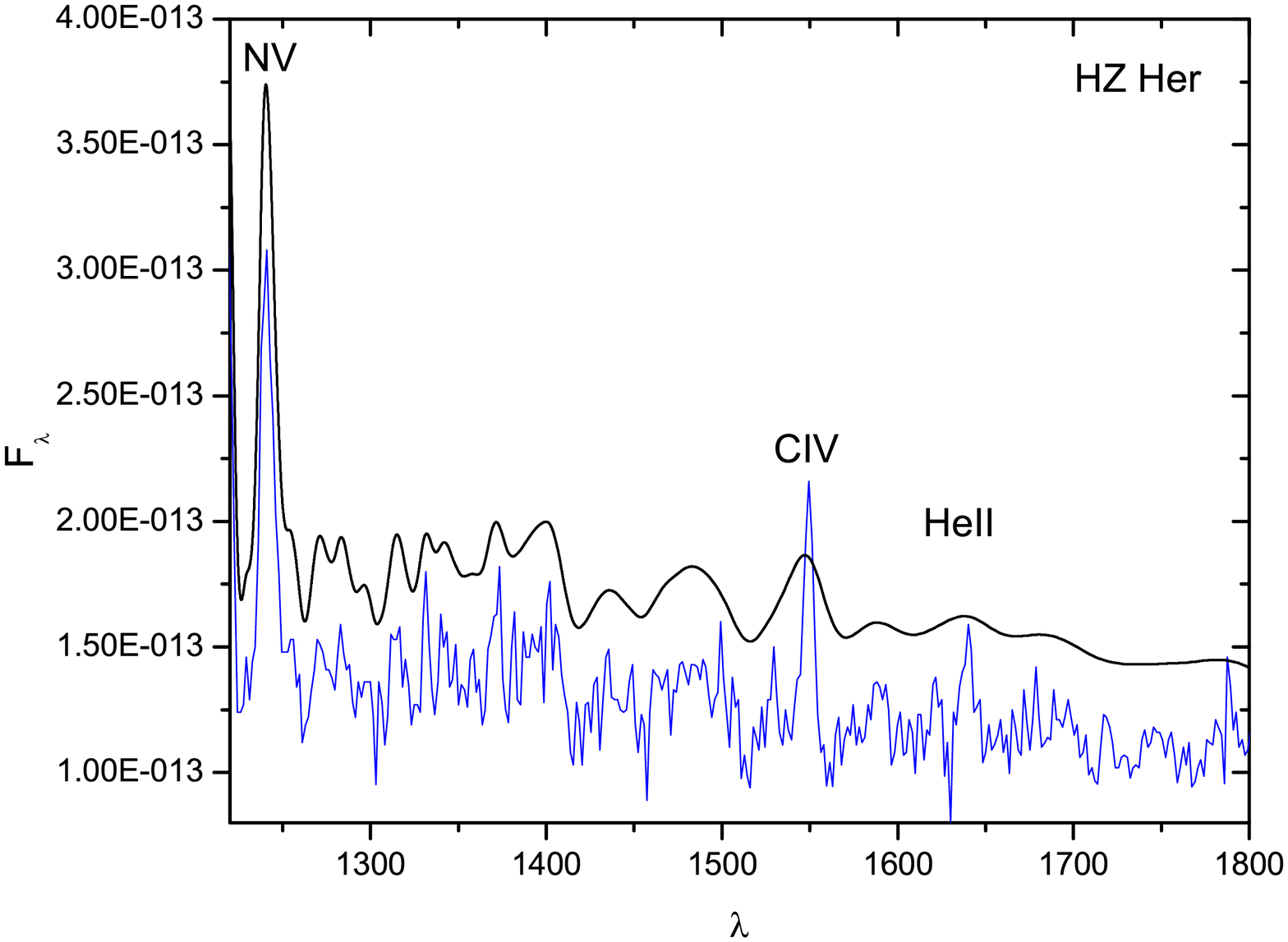}

\caption{The IUE spectra for (a) a HMXB 4U1700-37,  (b) a HMXB Cyg X-1, (c) an
IMXB LMC X-3, and (d) an IMXB Her X-1.  The smoothed curves are obtained by
convolving the IUE spectrum with the varying spectral resolution of the HST/ACS
prism. }

\end{figure}

Low Mass X-Ray Binaries: Typically, these have a neutron star with a companion
star of about 1M$_{\sun}$, with Sco X-1 or Cen X-4 being examples (Fig. Cen
X-4).  The strongest features are usually NV $\lambda$ 1240, Si IV $\lambda$
1400, C IV $\lambda$ 1550, and He II $\lambda$ 1640, all of which are visible
at the prism resolution with the exception of the He II line, which can appear
weak.  In Sco X-1, the C IV line is the strongest FUV feature (Fig. 4c). 

White Dwarf Systems: We mention these for completeness, as they are certainly
not responsible for the ULX systems but have relevant UV spectra.  The
cataclysmic binary SS Cyg has a white dwarf of mass 0.40 M$_{\sun}$ and a dwarf
star of 0.6 M$_{\sun}$, and its UV spectrum shows a very strong line of CIV
$\lambda$ 1550\AA, a somewhat weaker Si IV $\lambda$ 1400\AA\ feature, and
weaker lines NV $\lambda$ 1240\AA, He II $\lambda$ 1640\AA, and two lower
ionization lines of OI and C II (Fig. SSCyg).  Supersoft sources, which we
assume have a white dwarf as the compact object, such as RXJ 0513.9-6951, have
strong lines of He II 1640\AA\ and N V 1240\AA\ (Fig 4d).

\begin{figure}

\plottwo{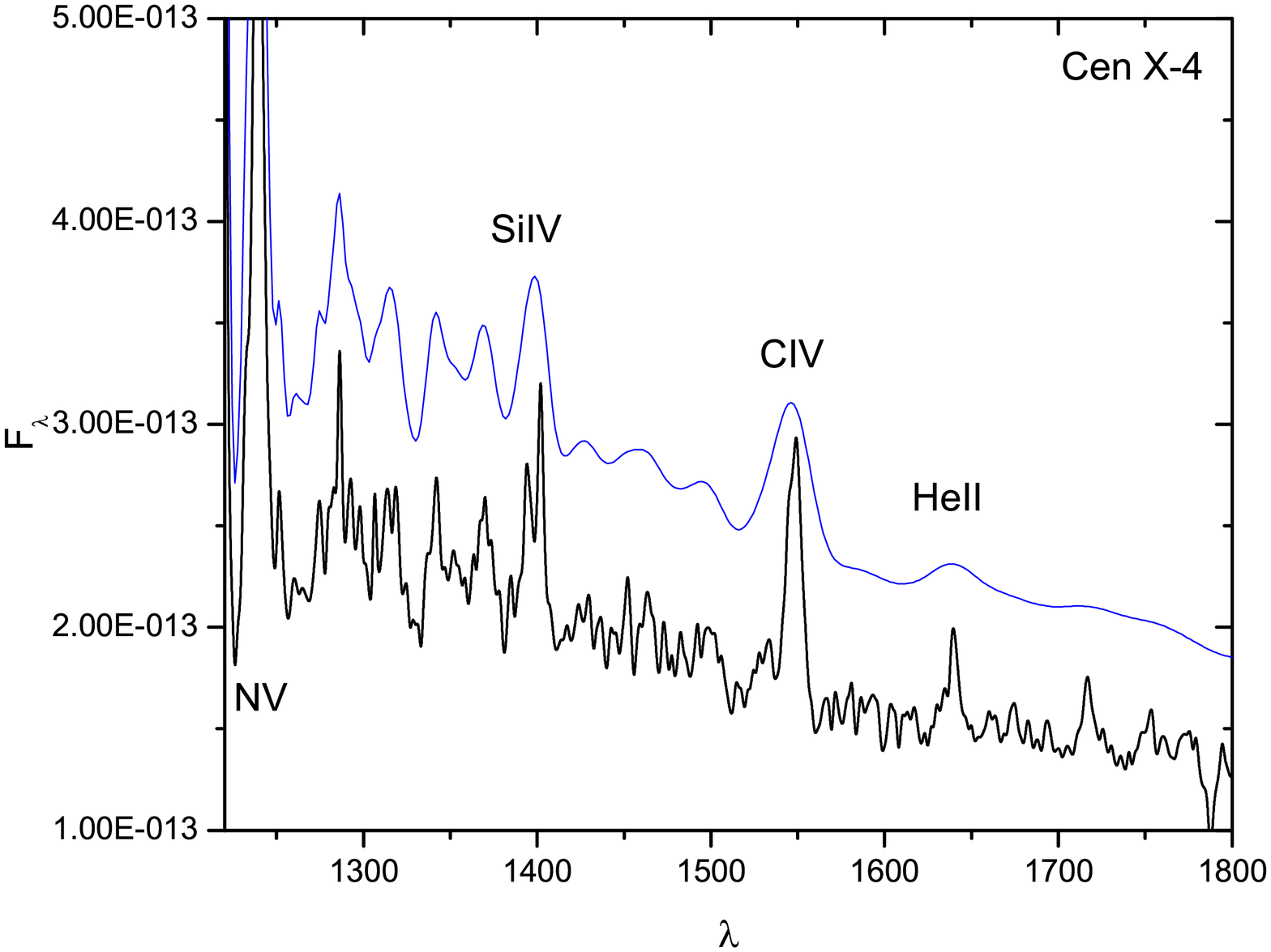}{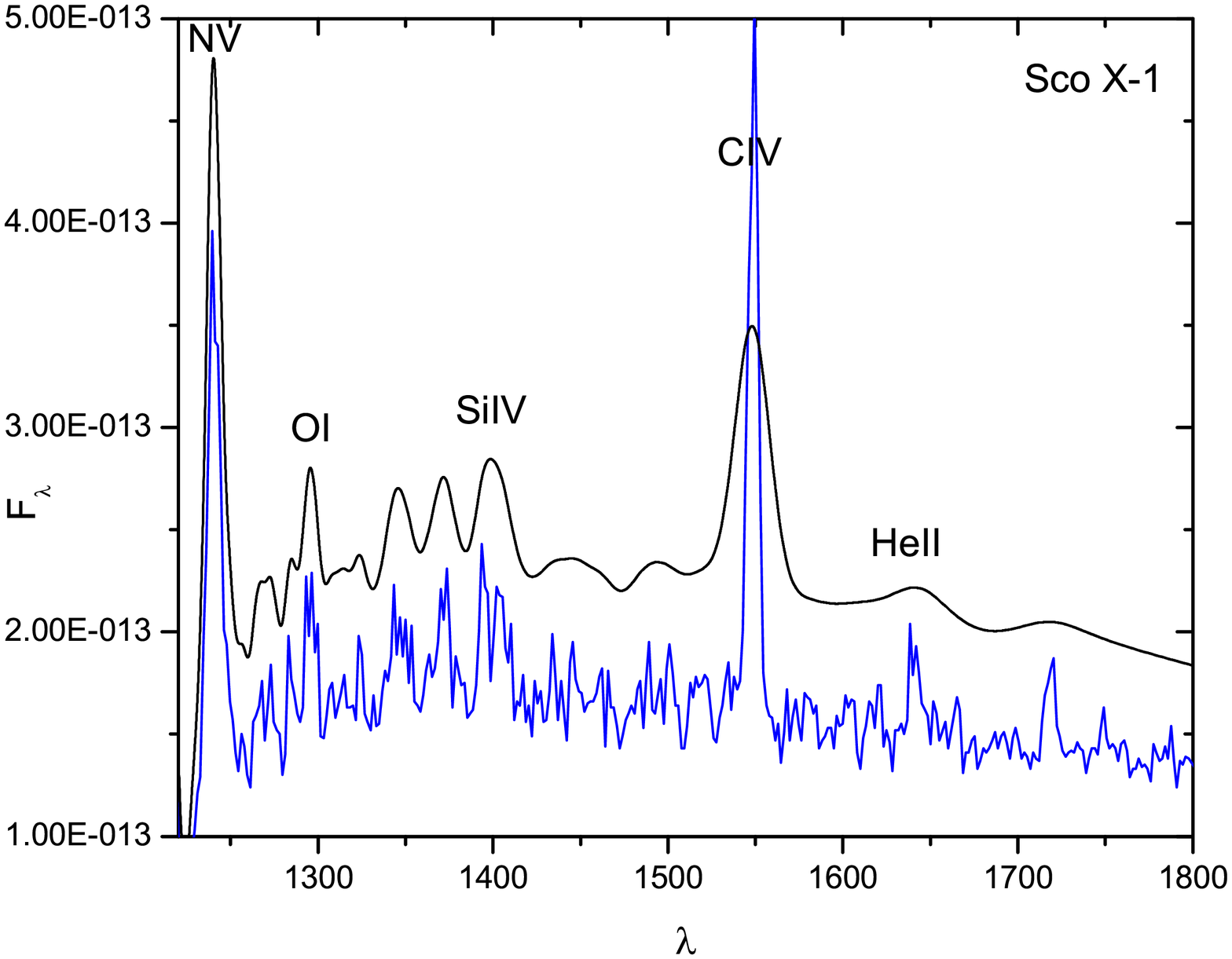}

\plottwo{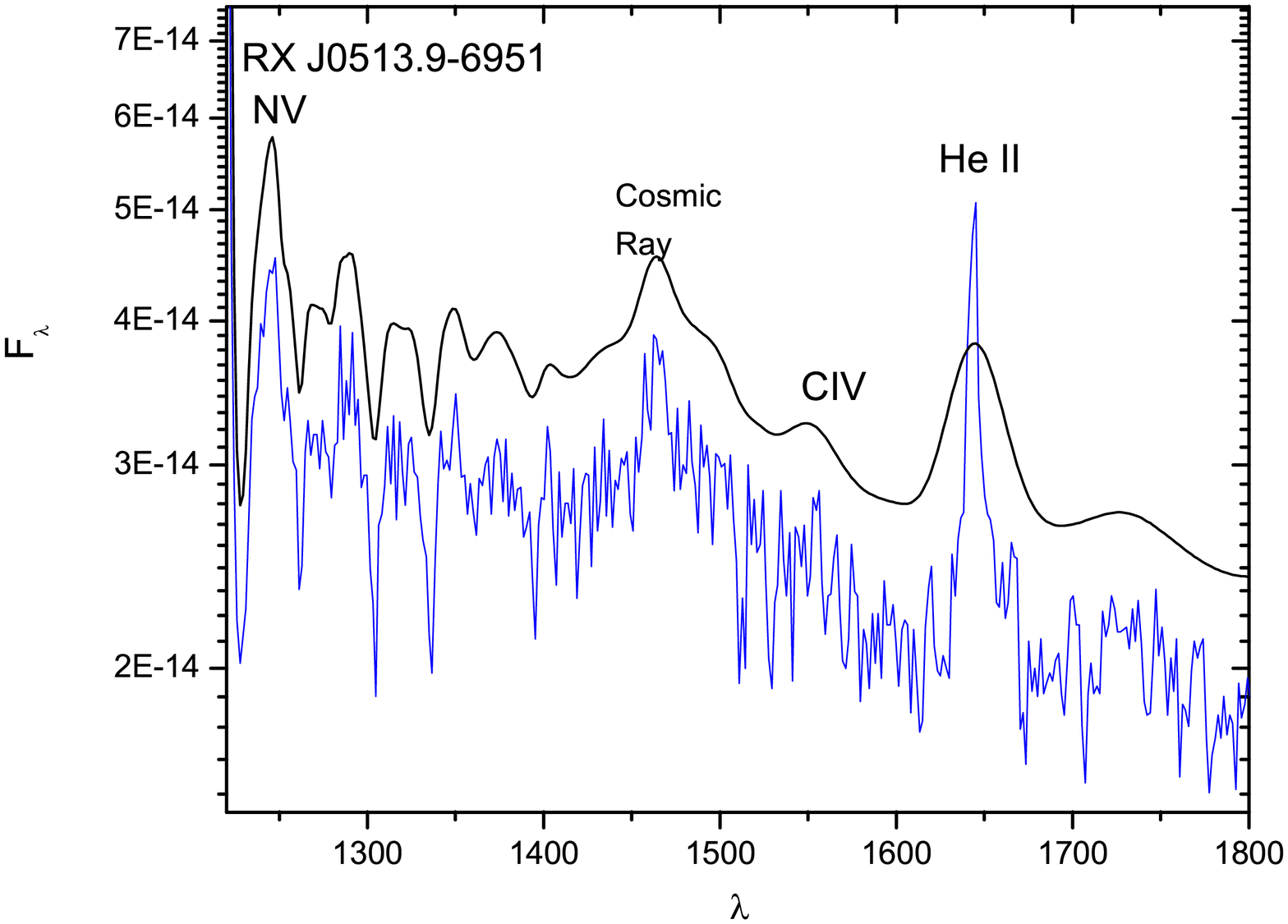}{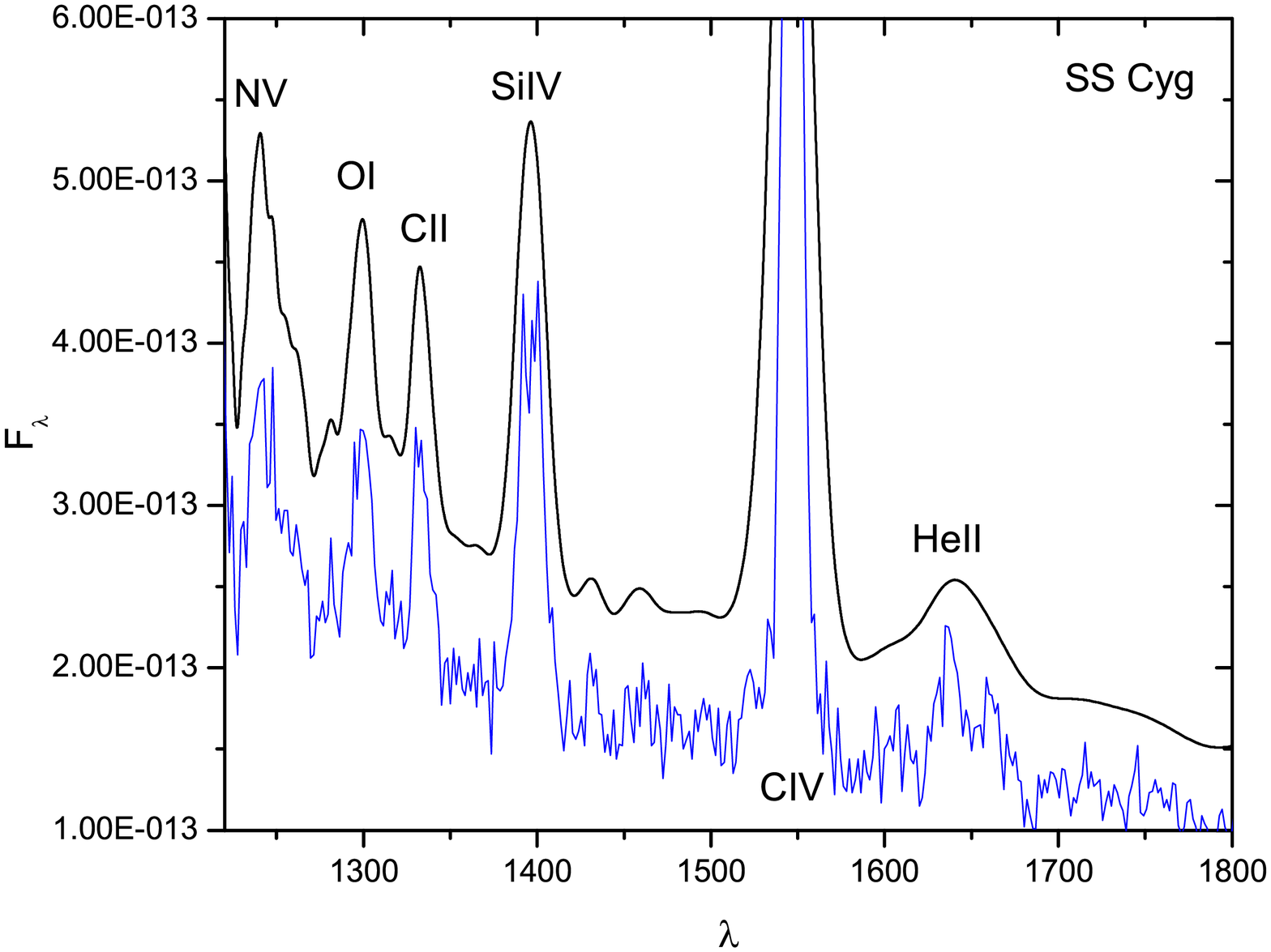}

\caption{The IUE spectra for (a) a LMXB Cen X-1,  (b) a LMXB Sco X-1, (c) a
supersoft source RX J0513.9-6951, and (d) a cataclysmic variable SS Cyg.  The
smoothed curves are obtained by convolving the IUE spectrum with the varying
spectral resolution of the HST/ACS prism. }

\end{figure}

\subsection{Comparison to the ULX Prism Spectra}

In comparing these spectra to the four ULX systems, we begin with Ho II ULX and
NGC 1313 ULX, which are quite similar to each other (Fig. 1a-b).  They both
show the N V $\lambda$ 1240 feature as their strongest emission line,
found in most classes of mass-transfer binaires.  These spectra indicate
that the two ULX candidates are not single O/B stars, thus confirming the
identification of the optical counterpart.  There is no strong absorption
feature at the location of the Si IV $\lambda$ 1400 or C IV $\lambda$ 1550
doublets, which are characteristic of HIMXBs and early-type WR stars.  This
would indicate that these two ULX systems are not HIMXBs in which the mass
transfer is dominated by winds.  The closest match of these two ULX spectra is
with intermediate mass XRBs, although the match is not perfect.  In HZ Her,
the equivalent width of the N V $\lambda$ 1240 line is nearly four times larger
than the ULXs in Ho II or NGC 1313, with a relative difference in the same
sense for the C IV $\lambda$ 1550 line (Table~1).  In contrast, the ULXs have
relatively stronger He II $\lambda$ 1640 lines than in HZ Her, but the
difference is only 40\%.  These relative equivalent width differences are also
true for a comparison to the LMXB Cen X-4, with the similarity to HZ Her
probably due to a common structure to the accretion disk, although the C IV
$\lambda$ 1550 line is stronger in Cen X-4.  The differences between the ULX
spectra and that of the HZ Her may be due to a lower metallicity in the donor
star for the ULX systems, it could be due to greater continuum dilution by the
ULX donor stars, or it could indicate a more significant difference in the
structure of the accretion disk.

The ULX in NGC 5204 is the only ULX source with a good-quality STIS low
resolution far UV spectrum, previously reported by Liu et al. (2004; Fig. 5).
The STIS G140L spectrum is of higher S/N than that from the prism, and we
compare the two by convolving the STIS spectrum to the resolution of the prism.
The N V $\lambda$ 1240 line is present at about the same EW as the previous two
ULX systems (note that the other ULX sources are more than twice as bright and
have higher S/N).  A few other weak emission sources are known from the STIS
spectrum (marked in figure), but emission from C IV $\lambda$ 1550 was not
detected while the He II $\lambda$ 1640 is formally present at about the 4-5
$\sigma$ level.  The convolved STIS spectrum is consistent with the prism
spectrum, showing the O I/SiII emission feature, preceded by absorption.  The
NGC 5204 ULX system is similar to the ULX systems in NGC 1313 and Ho II.  With
the STIS spectrum, stellar photospheric lines are detected, which allows for a
separation of disk and secondary, which is type B0 Ib (Liu et al. 2004), or
about 25 M$_\sun$ (Schmidt-Kaler 1982).  This secondary is more massive and
more luminous than the counterpart in HZ Her.  Consequently, it would be
expected that the stellar continuum is relatively more significant than the
lines, diminishing their equivalent widths.  Another possibly important feature
is the presence of the N IV] $\lambda$ 1486 line, seen tentatively in each ULX
prism spectrum but detected above the 5$\sigma$ level in the STIS spectrum of
NGC 5204.  This semi-forbidden line is not detectable in a stellar photosphere
because the density is above the critical density.  It is not predicted in flat
accretion disks with atmospheres (Raymond 1993), nor is it detected in LMXB or
CV systems, but it is detected in some IMXB systems, such as LMX X-3.  For this
line to occur, there must be an adequate emission measure of gas that is
strongly excited, which can occur in outflows, such as in Wolf-Rayet stars, but
this is not the only possibility.

\begin{figure}

\plotone{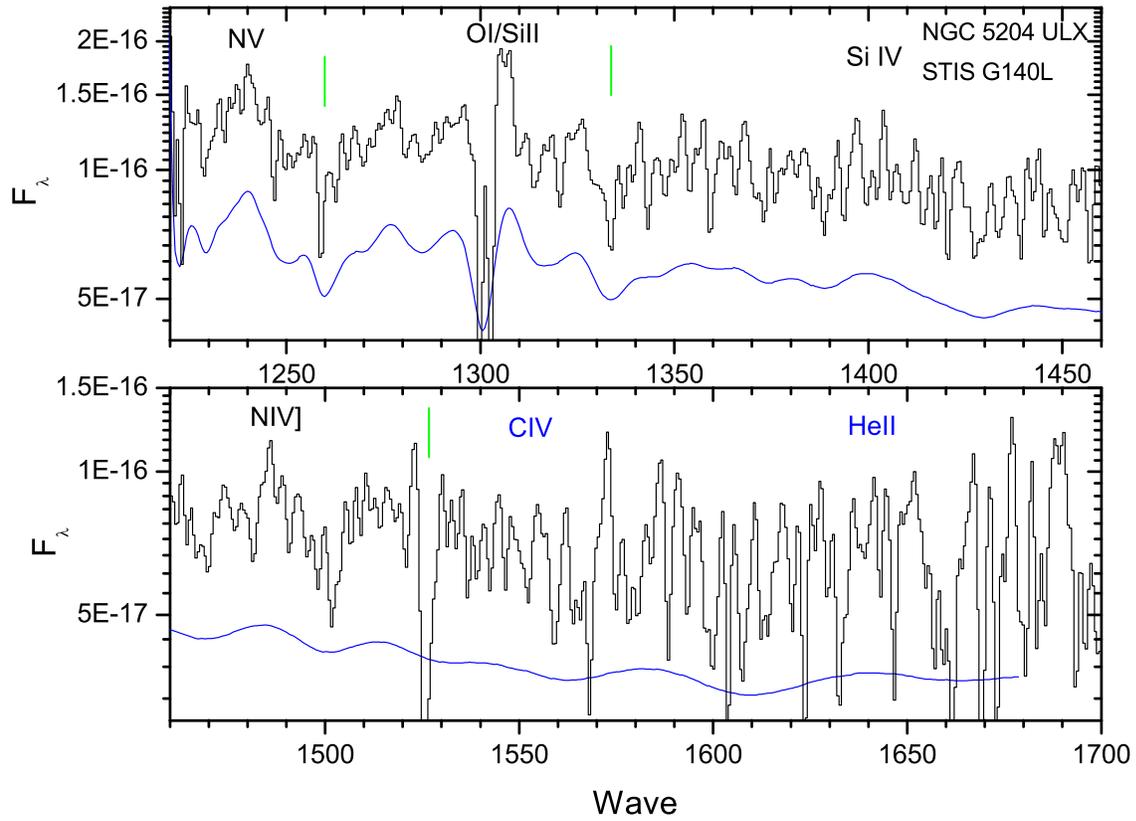}

\caption{The HST/STIS FUV spectrum for the ULX in NGC5204. The smoothed 
curve is obtained by convolving the IUE spectrum with the varying
spectral resolution of the HST/ACS prism. }

\end{figure}

The flux density of the ULX in M81 is about 30 times lower than for the ULXs in
NGC 1313 or Ho II, so the quality of the features is poor (Fig. 1d). The N V
$\lambda 1240$ emission line is not detected given the low S/N at the blue end,
although there is an indication of emission near this wavelength.  Three
possible emission features in the spectrum as labeled in Fig. 1d would appear
to be most similar to the late-type Wolf-Rayet star, HD86161 (Fig. 2d), although
better quality data is needed to examine this possibility.

\subsection{Color-Color Plots}

In addition to using the spectral features of ULXs and comparison sources, we
can also use the shape of the continuum.  The reason for using this is that
there are different expectations for the shape of the continuum from a
multi-temperature disk and from a hot star.  Thus, it may be possible to
distinguish between a continuum dominated by an accretion disk and a continuum
dominated by hot stars.

There are a few challenges facing this approach, such as choosing the location
of the continuum points and compensating for extinction along the line of
sight.  For the continuum points, we chose locations that are not dominated by
emission lines yet provide the best leverage for a two-color analysis: 1300\AA;
1500\AA; and 1800\AA.  The second issue is compensating for the extinction, but
the reddening vector can be determined from the extinction law of Cardelli et
al. (1992).  The ULXs are known to generally occur in regions of extinction
where the intervening column is typically 1-5$\times 10^{21}$ cm$^{-2}$,
although the metallicity is often sub-solar.  Extinction usually affects hot
stars, which occur near their dust-rich birth sites.  Also, many of the XRBs
lie in the disk of the Milky Way, where extinction can be considerable.

In our UV color-color analysis, we also show the color expected from a standard
multi-temperature disk, where $ F_\nu \propto \nu^{1/3}$, as well as the
Rayleigh-Jeans limit ($F_\nu \propto \nu^2$) to the blackbody curve.  A very
hot star would approach this Rayleigh-Jeans limit, but as the star becomes
progressively cooler, it approaches the multi-temperature disk line for a
blackbody temperature of about 2.5$\times 10^4$ K in the chosen color bands as
shown in Figure~6.  The XRBs of various types scatter around the
multi-temperature disk line.  The heavily reddened Cygnus X-1 lies on this
relationship, indicating that the choice of the reddening law leads to sensible
results.  The stars, which are mostly hot, are distributed from the
Rayleigh-Jeans line to the multi-temperature disk line.  The Wolf-Rayet stars
indicate significant extinction, as would be expected for stars with a large
amount of mass loss and embedded in star forming regions, while many of the
other hot stars were chosen because of their more modest absorption, and they
lie lower in the diagram.  The ULX sources lie between the multi-temperature
disk model and the Rayleigh-Jeans model, somewhat more to the Rayleigh-Jeans
side of the XRBs.  This is consistent with a model where the ULX sources are a
combination of a stellar continuum and an accretion disk source.  From the
spectral analysis, we concluded that the ULX in NGC 5204 has a higher ratio of
stellar continuum and that the ULXs in NGC 1313 and Ho II have relatively more
accretion disk emission.  Their location on the color-color plot is consistent
with that interpretation, as those latter two ULXs have colors closer to the
multi-temperature disk line.  We view this approach as a useful consistency
check for our interpretation of the ULX spectra.

\begin{figure}

\plotone{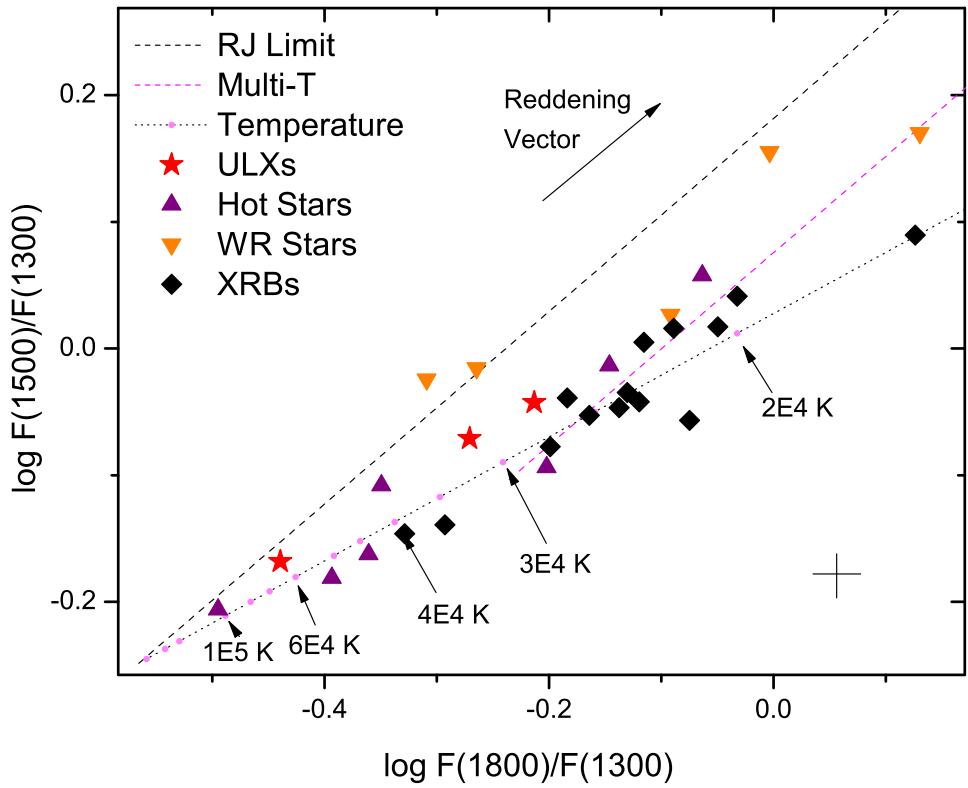}

\caption{Color-color diagram for the four ULXs and other known objects.
Overplotted curves include  the reddening vector, the standard
multi-temperature disk and the Rayleigh-Jeans limit with varying extinction,
and the black body models with different temperatures. The S/N of the ULX in
M81 was too poor to provide useful color data, so that object is not included.
}

\end{figure}

\section{Discussion}

One of the primary goals of this work was to determine whether the optical
counterparts, identified by finding an object within an error circle, was
indeed the donor star in the ULX binary system.  For the ULX in NGC 5204, this
was already established from a STIS UV spectrum (Liu et al. 2004), which showed
the N V $\lambda$ 1240 emission feature, a feature predicted to be strong for
accretion disk systems (Raymond 1993).   For the ULX optical counterparts in
NGC 1313 and Ho II, there were significantly stronger UV emission line
features, which may be due to a larger contribution from the accretion disk
relative to the star. For the ULX in M81, this line is not detected due to the
low S/N at the blue end, but there is an indication of emission near this line.

The stellar mass can be estimated from the magnitude and color with the help of
the high resolution UV spectrum.  In the case of NGC5204, the high resolution
STIS spectrum was available, which showed stellar photospheric lines and
allowed for a separation of the disk and a B0Ib secondary of about 25$M_\odot$
(Liu et al. 2004).  No such mass estimates are made for the ULXs in Ho II and
NGC1313 because there is a degeneracy with spectral type due to the unknown
fraction of emission coming from the accretion disk. This degeneracy can be
removed with a higher resolution UV spectrum, which is possible with current
instruments on HST.

The secondary mass for the ULX in Ho II has been estimated in previous works
with the help of emission lines in other bands.  Strong He II $\lambda4686$
emission line was detected in the surrounding nebula produced by X-ray
photoionization, and the secondary is estimated to be in the range of B3Ib to
O4V based on the optical data (Kaaret et al. 2004).  With Spitzer, Berghea et
al. (2010) detected the [OIV] 25.89 $\mu$m line, and estimated the secondary to
be a B2Ib supergiant based on the overall spectral energy distribution from
X-ray to UV to infrared. Such a secondary is about 20$M_\odot$, similar to the
secondary in NGC 5204.

For the ULX in NGC1313, the secondary mass is estimated from $\sim8.5M_\odot$
(Liu et al. 2007) to $\sim20M_\odot$ (Mucciarelli et al. 2005) based on the
optical magnitudes and different separation of the secondary and disk
contribution. Recently, Liu et al. (2012) estimated the secondary to be a
$\sim8M_\odot$ evolved star based on the dynamical constraints by combining the
HST light curves, including the orbital period of $\sim6$ days (Liu et al. 2009), and the
radial velocity amplitude measured from the He II $\lambda4686$ emmision line
from the disk (Roberts et al. 2011). In this model, the emission is dominated
by the accretion disk; this is consistent with the stronger N V $\lambda$ 1240
emission line as compared to the case of NGC 5204.

There is a similarity in the nature of the secondary for the ULX counterpart in
NGC 1313, Ho II, and NGC 5204 in that they are blue evolved stars of about
8--20 M$_{\sun}$.  By mass, this might suggest that they are high mass X-ray
binaries, in which mass transfer is often through stellar winds.  However,
these systems do not show in the UV P-Cygni like absorption features that
accompany HIMXBs.  Rather, these are similar to the intermediate mass binaries,
where the mass transfer is through Roche Lobe overflow.  If this type of
configuration is common in ULXs, it could indicate the binary properties
necessary for the ULX phenomenon to occur.

\acknowledgements

The authors gratefully acknowledge valuable commentary and assistance from a variety of
people, including Jon Miller, Jimmy Irwin, Sally Oey, Charles Cowley, Mark Reynolds, and Ed Cackett.  
We gratefully acknowledge financial support for this work from NASA.
Some of the data presented in this paper were obtained from the Mikulski Archive for 
Space Telescopes (MAST). STScI is operated by the Association of Universities for 
Research in Astronomy, Inc., under NASA contract NAS5-26555. Support for MAST for 
non-HST data is provided by the NASA Office of Space Science via grant NNX09AF08G and 
by other grants and contracts.

\begin{deluxetable}{llcccccc}
\tablecolumns{8}
\tablecaption{Emission Line Equivalent Widths}
\tablehead{
	\colhead{Binary Name} & \colhead{Type} & \colhead{NV} & \colhead{CIV} & \colhead{HeII}
	 & \colhead{F(NV)} & \colhead{F(He II)} & \colhead{Fx}
	 }
\startdata
Ho II 	& ULX	& 2.7 $\pm$ 0.3	& \textless\ 2	& 2.4 $\pm$ 0.4	& 6.1E-16 (6)& 4.2E-16 (5) & 1.2E-11\\
NGC 1313 & ULX	& 2.7 $\pm$ 0.3	& \textless\ 2 & 2.4 $\pm$ 0.4 & 5.2E-16 (6) & 3.8E-16 (5) &	5E-12\\
NGC 5204 & ULX	& 2.9 $\pm$ 0.4	&  \textless\ 2 & 2.0 $\pm$ 0.4 & 2.8E-16 (5) & 1.2E-16 (3) &	2E-12\\
HZ Her & IMXB	& 10. $\pm$ 1. & 3.2 $\pm$ 0.3 & 1.7 $\pm$ 0.3 & & & \\
Cen X-4 & LMXB	& 11. $\pm$ 1. & 7.0 $\pm$ 0.7 & 1.5 $\pm$ 0.2 & & & \\
\enddata
\end{deluxetable}


\begin{thebibliography}

\bibitem[Begelman (2002)]{} Begelman, M.C., 2002, \apj, 568, L97
\bibitem[Berghea et al. (2010)]{} Berghea, C. T.; Dudik, R. P.; Weaver, K. A. \& Kallman, T. R., 2010, ApJ, 708, 354
\bibitem[Cardelli et al. (1992)]{} Cardelli, J; Sembach, K. \& Mathis, J., 1992, AJ, 104, 1916
\bibitem[Colbert et al. (1999)]{} Colbert, E. J. M. and Mushotzky, R. F. 1999, \apj, 519, 89
\bibitem[King et al. (2001)]{} King, A. R., Davies, M. B., Ward, M. J., Fabbiano, G. and Elvis, M.  2001, \apj, 552, L109
\bibitem{} K\"{u}mmel, M., Walsh, J.R., Pirzkal, N. et al. 2009, PASP, 121, 59
\bibitem[Liu \& Bregman (2012)]{} Liu, J. \& Bregman, J., 2012, ApJS, submitted
\bibitem[Liu et al. (2002)]{} Liu, J., Bregman, J., and Seitzer, P., 2002, ApJL, 580, 31
\bibitem[Liu et al. (2004)]{} Liu, J., Bregman, J., and Seitzer, P., 2004, ApJ, 602, 249
\bibitem[Liu et al. (2007)]{} Liu, J.; Bregman, J.; Miller, J. \& Kaaret, P., 2007, ApJ, 661, 165
\bibitem[Liu et al. (2009)]{} Liu J., Bregman J., \& McClintock, J., 2009, ApJL, 679, 37
\bibitem[Liu et al. (2012)]{} Liu, J., Orosz, J., \& Bregman, J., ApJ, 2012, 745, 89
\bibitem[Mucciarelli et al. (2005)]{} Mucciarelli, P.,Zampieri, L., et al. 2005, ApJ, 633, L101
\bibitem[Niedzielski \& Rochowicz (1994)]{} Niedzielski, A. \& Rochowicz, K., 1994, A\&AS, 108, 669
\bibitem[Prinja (1990)]{} Prinja, R.K., 1990, MNRAS, 246, 392
\bibitem[Raymond (1993)]{} Raymond, J. 1993, ApJ, 412, 267 
\bibitem[Roberts et al. (2011)]{} Roberts, T.~P., Gladstone, J.C., Goulding, A.D., et al. 2011, AN, 332, 398
\bibitem{} Schmidt-Kaler, Th., 1982, {\it Landolt-Bornstein: Numerical Data and Functional Relationships in Science and Technology}, eds. Schaifers, K. and Voigt, H.H., VI/2b, P.15-31


\end{thebibliography}
\end{document}